\documentclass[12pt]{article}%
\usepackage{amsmath,amssymb,amsfonts}
\usepackage{graphicx,epsfig}
\usepackage{hyperref}
\usepackage{color}
\usepackage{amsmath}
\usepackage{amsfonts}
\usepackage{graphicx}
\usepackage{slashed}%

\newcommand{\NN}{\mbox{${\mathbb N}$}}

\def\ri{{\rm i}}

\def\k{\kappa}

\numberwithin{equation}{section}
\begin{document}
\begin{titlepage}
\begin{flushright}
\end{flushright}
\vskip 1.0cm
\begin{center}
{\Large \bf Dualities for anyons} \vskip 1.5cm
{\large Brando Bellazzini}\\[0.5cm]
{\it Institute for High Energy Phenomenology\\
Newman Laboratory of Elementary Particle Physics\\
Cornell University, Ithaca, NY 14853, USA} \\[5mm]
\vskip 2.0cm 
\abstract{We show that the low-energy dynamics of  anyons in (1+1)-dimensions with the smallest number of derivatives and  $\mathcal{C}$, $\mathcal{P}$ and $\mathcal{T}$ symmetric interactions, are dual to the sine-Gordon model for bosonic fields. We discuss in particular the Tomonaga-Luttinger, Thirring and Schwinger models,  as well as their deformation by relevant and marginal operators. 
In the presence of electromagnetic interactions, the mass of the meson from anyon confinement and the chiral anomaly get corrected by the statistical parameter.}
\end{center}
\end{titlepage}


\section{Introduction} 
In three spatial dimensions integer-spin particles are bosons  with symmetric wave-functions under permutations, while half-integer spins are fermions which are antisymmetric under the exchange of quantum numbers.  
Conversely, in lower dimensions generalized anyonic statistics which interpolate between bosons and fermions, are possible \cite{Leinaas:1977fm}. 
In two spatial dimensions, the rotation group is abelian, the spin can be an arbitrary real number, and it is well known that  fractional braiding statistics describe elementary excitations in the quantum Hall effect.
In one spatial dimension (1D) there is no rotation that can move particles around each other and the only way to exchange them is through collisions which eventually relate statistics and interactions. Furthermore, anyonic statistics in 1D might emerge from 
non equivalent self-adjoint  extensions of the kinetic term \cite{Hansson:1991uc}, corresponding to two-dimensional anyons restricted to the first Landau level by a strong magnetic field.

Attention to 1D systems is mainly motivated by  quantum Hall fluids where transport is localized on the edge and it is due to 1D chiral anyons. 
Recent experimental realizations of trapped 1D atomic gases  \cite{1dexp} and the possibility of engineering an anyonic gas in rapidly rotating trap \cite{par} has led to renewed theoretical interest in 1D anyonic models 
\cite{k-99,cm-07,sc,ssc-07,Bellazzini:2008fu,lm-99,it-99,pka,bgk,oae-99,%
g-06,an-07,kl-05,g-07,zw-07,o-07,hzc-08,dc-08,LeClair:2004bd,Engquist:2008mc,Kundu:2010xb}.

Motivated by the desire to capture the general and model independent properties of anyons in 1D, we focus on the low-energy description of generic  anyonic interactions as dictated by symmetries and renormalization. 
We classify all possible renormalizable (self-)interactions according to their properties under charge-conjugation $\mathcal{C}$, parity $\mathcal{P}$, and time-reversal symmetry $\mathcal{T}$. In particular, we show that the most general 4-anyon interaction symmetric under $\mathcal{C}$, $\mathcal{P}$, and  $\mathcal{T}$  is equivalent by means of bosonization to the sine-Gordon model for a bosonic field.  A Lorentz symmetry that preserves either the light-cone or the sound-cone emerges at low energy as an accidental symmetry.  Also, we show that $U(1)$ gauge interactions between photons and anyons (that give rise to anyon confinement) are dual to the massive sine-Gordon model.
These results extend the well known dualities among fermionic systems and the sine-Gordon model \cite{Coleman:1974bu,Schwinger:1962tp,Lowenstein:1971fc,Kogut:1974kt,Coleman:1975pw,Coleman:1976uz,Mandelstam:1975hb}
 to anyons with generic renormalizable interactions. 

The paper is organized as follows. In the next section we introduce free anyon models with the smallest number of derivatives and we discuss the symmetry content of the theory. In section \ref{bosonization} we use bosonization to formulate anyons and their currents in terms of bosonic variables. In section \ref{sect-duality} we discuss three solvable models (the anyonic Tomonaga-Luttinger, Thirring, and Schwinger models) and the impact of the most general renormalizable deformation that respect $\mathcal{C}$, $\mathcal{P}$, and $\mathcal{T}$. We also discuss the modification of the chiral anomaly when the anyons are electrically charged and derive the mass of the composite state from anyon confinement. Section \ref{sect-conclusions} is devoted to our conclusions. 
Appendix \ref{deform} contains the classification of all renormalizable anyonic interactions. Appendix \ref{app-kcomm} contains a discussion of the $\k$-commutator for massless anyons.

\section{Low-energy anyons}

Simple scaling arguments and power counting suggest that the low-energy behavior of any field theory is captured by an effective lagrangian with the smallest number of fields and derivatives. 
Thus, we look for the low-energy anyonic excitations encoded into a renormalizable lagrangian containing derivatives up to   first order.

 An anyonic field $\psi$ satisfies exchange relations at $x_{1}\neq x_{2}$ controlled by the statistical parameter $\k$
 \begin{align}
 \label{exch1}
\psi(t,x_1)\psi(t,x_2)=&\psi(t,x_2)\psi(t,x_1)e^{-i\pi\k\varepsilon(x_{1}-x_{2})}\\
 \label{exch2}
\psi^*(t,x_1)\psi(t,x_2)=&\psi(t,x_2)\psi^*(t,x_1)e^{i\pi\k\varepsilon(x_{1}-x_{2})}
\end{align}
where $\varepsilon(x)$ is the sign function\footnote{$\varepsilon(x)=-\varepsilon(-x)=1$ for $x>0$ and $\varepsilon(0)=0$.}.
Odd (even) integer values of $\k$ correspond to fermions (bosons). Non-integer values are also possible in 1D and give  rise to general anyonic statistics.  The exchange relation between  $\psi$ and $\psi^*$ is singular at coincident points and it is discussed
 in appendix \ref{app-kcomm}.
  
For a gapless anyon, the simplest $\mathcal{T}$-symmetric equations of motion with only first-order derivatives
requires two anyons $\psi_{1}$ and $\psi_{2}$ which describe free left- and right-movers  traveling at the Fermi velocity $v_{F}$
\begin{align}
\label{eom1}
\left(\partial_t+v_{F}\partial_{x}\right)\psi_{1}=0\,,\quad \left(\partial_t-v_{F}\partial_{x}\right)\psi_{2}=0\,.
\end{align}
The $\mathcal{T}$ symmetry imposes that $\psi_{1}$ and $\psi_{2}$ have opposite statistical parameters, $\k$ and $-\k$ respectively.
Equations (\ref{eom1}) turn out to be symmetric under $\mathcal{C}$ and $\mathcal{P}$  discrete symmetries as well.  
Of course, (\ref{eom1}) are also covariant under a continuous Lorentz symmetry 
\begin{equation}
\label{boost0}
\psi_{\alpha}(t,x)\rightarrow e^{-(-1)^{\alpha}\chi \k/2}\psi_{\alpha}( \Lambda_{\chi}(t,x))
\end{equation}
 where  $\Lambda_{\chi}$ is a boost, labeled by the rapidity $\chi$, that leaves the\textit{ light-cone} $(v^2_{F}t^2-x^2)=0$ invariant
 \begin{equation}
 \label{boost00}
 \Lambda_{\chi}(t,x)=(t^\prime,x^{\prime})=\gamma(t-\frac{\beta}{v_{F}} x,x-\beta v_{F} t)\,,\qquad \gamma=(1-\beta^2)^{-1/2}\,,\qquad \beta=\tanh\chi\,.
 \end{equation}
 We set $v_{F}=1$ hereafter.
A global $U(1)_{V}\times U(1)_{A}$ chiral symmetry
\begin{align}
\label{vectorial}
U(1)_{V}:&\,\,\psi_{1}\rightarrow  e^{\ri \omega_{V}}\psi_{1}\qquad \psi_{2}\rightarrow  e^{\ri \omega_{V}}\psi_{2}\,,\\
\label{axial}
U(1)_{A}:&\,\,\psi_{1}\rightarrow  e^{\ri \omega_{A}}\psi_{1}\qquad \psi_{2}\rightarrow  e^{-\ri \omega_{A}}\psi_{2}\,,
\end{align} 
leaves (\ref{eom1}) invariant too. The $U(1)_{V}$ is identified with electromagnetism, left unbroken throughout the paper.

It is well known that a free Dirac particle in 1D admits an equivalent description in terms of free bosons. 
This remains true if one demands anyonic exchange relations as we will discuss in Section \ref{bosonization}. 
In particular,  (\ref{eom1}) admit a local action description only in terms of those bosonic fields.
In this paper we deform that maximally symmetric  action  by adding symmetry-breaking renormalizable terms $\mathcal{O}$  (i.e. with canonical dimension $[\mathcal{O}]\leq 2$)   to the lagrangian,  $\delta\mathcal{L}=\sum_{\mathcal{O}}c_{\mathcal{O}}\mathcal{O}$.
In App. \ref{deform} we show that there are only four relevant or marginal deformations $\mathcal{O}$ 
that respect $U(1)_{V}$ and are symmetric
 under $\mathcal{C}$, $\mathcal{P}$, and  $\mathcal{T}$: a mass term and four 4-anyon interactions
\begin{align}
\label{Odeform}
\bar{\psi}\psi\,,\quad (\bar{\psi}{\psi})^{2}\,,\quad (\bar{\psi}\gamma^\mu\psi)(\bar{\psi}\gamma_\mu\psi)\,,\quad \rho_{\pm}^2
\end{align} 
 where $\psi=(\psi_{1},\psi_{2})^{T}$, $\bar{\psi}=\psi^{*}\gamma^0$, $\gamma^0=\sigma_{1}$, and $\gamma^{1}=-\ri\sigma_2$. $\sigma_{1,2}$ are the first two Pauli matrices. In particular, $\rho_{\pm}=\psi^*_{1}\psi_{1}\pm \psi^*_{2}\psi_{2}$ generate $U(1)_{V}$ and $U(1)_{A}$ 
\begin{align}
\label{WardV}
[\rho_{+}(t,x),\psi_{\alpha}(t,y)]=&-\psi_{\alpha}\delta(x-y)\\
\label{WardA}
[\rho_{-}(t,x),\psi_{\alpha}(t,y)]=&(-1)^{\alpha}\psi_{\alpha}\delta(x-y)\,.
\end{align}
The operator $\bar{\psi}\psi=\psi^*_{1}\psi_{2}+\psi^*_{2}\psi_{1}$ couples different chiralities and it corresponds to a mass deformation.
Among the operators in (\ref{Odeform}), it is clear that the Tomonaga-Luttinger operator $(g_{+}\rho_{+}^2+g_{-}\rho_{-}^2)$ breaks Lorentz symmetry. However, we will see later that this breaking is very special and, in fact, we recover a Lorentz symmetry with respect to the sound wave velocity $v$ (i.e. that leaves the \textit{sound-cone} $(v^2t^2-x^2)=0$  invariant) where $v$ is fixed by coupling constants $g_{\pm}$.

In the next section we translate the anyon dynamics and the composite operators in (\ref{Odeform}) in terms of free bosonic fields.

\section{Bosonization and anyons}
\label{bosonization}

\subsection{Anyons from bosons}

Bosonization is the basic tool to study 1D anyonic interactions in terms of a lagrangian involving 
only bosonic degrees of freedom.  We follow the constructive operators approach used by Mandelstam \cite{Mandelstam:1975hb}.
We introduce  two free massless 
scalar bosonic fields $\phi$ and $\tilde{\phi}$ that are related by Hodge duality $\epsilon_{\mu\nu}\partial^\nu\phi=\partial_\mu\tilde\phi$, i.e.
\begin{equation}
\label{hodge}
\partial_{t}\phi=-\partial_{x}\tilde{\phi}\,,\quad \partial_{x}\phi=-\partial_{t}\tilde{\phi}\,.
\end{equation} 
This equation implies that both $\phi$ and $\tilde\phi$ satisfy the free massless Klein-Gordon equation.
Then, taking the usual commutation relations for scalar fields we get
\begin{align}
\nonumber
\phi(t,x)=&\int_{-\infty}^{\infty}\frac{dk}{2\pi\sqrt{2|k|}}\left\{a(k)e^{-i|k|t+ikx}+h.c.\right\}\\
\nonumber
\tilde{\phi}(t,x)=&\int_{-\infty}^{\infty}\frac{dk\, \varepsilon(k)}{2\pi\sqrt{2|k|}}\left\{a(k)e^{-i|k|t+ikx}+h.c.\right\}
\end{align}
with $[a(k),a^*(p)]=2\pi\delta(k-p)$.
Note also that any constant shift, 
\begin{align}
\label{shift}
\phi\rightarrow\phi+c\,,\quad \tilde{\phi}\rightarrow\tilde{\phi}+\tilde{c}\,,
\end{align}
leaves (\ref{hodge}) invariant. These symmetries are generated by generators $Q$ and $\tilde{Q}$
that commute with each other and give\footnote{A possible choice is given by $Q=\frac{1}{2}\int_{-\infty}^{+\infty}dx\partial_{t}\phi(t,x)$ and similarly for $\tilde{Q}$. 
For a rigorous definition of these charges see Ref. \cite{lm-99}.} $[Q,\phi(t,x)]=[\tilde{Q},\tilde{\phi}(t,x)]=-\ri/2$.

While both $\phi$ and $\tilde{\phi}$ are local, they are not relatively local, i.e. they don't commute at spacelike distances.
For instance, at equal times, 
\begin{equation}
\label{nonlocPh}
[\phi(t,x_1),\tilde{\phi}(t,x_2)]=\frac{\ri}{2}\varepsilon(x_{1}-x_{2})\,.
\end{equation}
This non locality between $\phi$ and $\tilde{\phi}$ is the crucial ingredient that eventually allows one to get anyons out of bosons. Indeed, for any pair of real numbers $\zeta_{\pm}$, we can define the anyon field $\psi_{i}$ by taking the exponentials of linear combinations of $\phi$ and $\tilde{\phi}$
\begin{align}
\label{psi1}
\psi_1(t,x)=& \eta :\mbox{Exp}\left\{\ri\sqrt{\pi}\left[\zeta_{+} \phi(vt,x) - \zeta_{-}\tilde{\phi}(vt,x)\right]\right\}:\\
\label{psi2}
\psi_2(t,x)=& \tilde{\eta} :\mbox{Exp}\left\{\ri\sqrt{\pi}\left[\zeta_{+} \phi(vt,x) + \zeta_{-}\tilde{\phi}(vt,x)\right]\right\}:
\end{align}
where $\eta$ and $\tilde{\eta}$ are constant operators (Klein factors) expressed in terms of exponential of the charges,
\begin{align}
\nonumber
\eta = & \frac{z}{\sqrt{2\pi}} :\mbox{Exp}\left\{\ri\sqrt{\pi}\left[ \zeta_{+} \tilde{Q}+\zeta_{-} Q \right]\right\}:\\
\nonumber
\tilde{\eta} = &  \frac{z}{\sqrt{2\pi}} :\mbox{Exp}\left\{\ri\sqrt{\pi}\left[ \zeta_{+} \tilde{Q}-\zeta_{-} Q \right]\right\}:\,,
\end{align}
and $:\ldots:$ represents normal ordering with respect to $a(k)$ and $a^*(k)$. The overall constant $z$ that fixes the normalization (and the dimension) is determined later.
We introduce the sound speed $v$ into the definition of $\psi_{\alpha}$ for future convenience, since we expect it to be renormalized in presence of non trivial interactions. From these definitions, we get that $\psi_{1}$ and $\psi_{2}$ satisfy anyonic exchange relations (\ref{exch1}, \ref{exch2}), controlled respectively by the statistical parameters $\k$ and $-\k$ expressed in terms of $\zeta_{\pm}$
\begin{equation}
\label{thetaZetas}
\k=-\zeta_{+}\zeta_{-}=\left\{
\begin{array}{ccc}
\mbox{odd} &\rightarrow & \mbox{ fermion}\\
\mbox{even} &\rightarrow & \mbox{boson}\\
\mbox{otherwise} &\rightarrow &\mbox{anyon}
\end{array}\right.\,.
\end{equation}

As a basic example, let us consider free massless anyons as described by (\ref{eom1}).
In this case it is clear that the equations of motion fix only $\zeta_{-}=-\zeta_{+}$, while generic interactions fix them as functions of the coupling constants (we will show exactly solvable examples in the next section).  Once $\zeta_{\pm}$ are given, the correlation functions of the interacting theory can be extracted using the identity (in Fock representation)
\begin{align}
\nonumber
:e^{A}:\,:e^{B}:=e^{\langle A B\rangle} :e^{A+B}: 
\end{align}
valid when $A$ and $B$ are linear combinations of $\phi$ and $\tilde{\phi}$. Thus, the basic correlators needed are 
\begin{align}
\label{bosoncorr1}
&\langle\phi(t_{1},x_{1})\phi(t_{2},x_{2})\rangle=  \langle\tilde{\phi}(t_{1},x_{1})\tilde{\phi}(t_{2},x_{2})\rangle \\
\nonumber
&=-\frac{1}{4\pi}\left\{\ln[\ri\mu(t_{12}-x_{12})+\epsilon]+\ln[\ri\mu(t_{12}+x_{12})+\epsilon]\right\}\\
\label{bosoncorr2}
&\langle\phi(t_{1},x_{1})\tilde{\phi}(t_{2},x_{2})\rangle=\langle\tilde{\phi}(t_{1},x_{1})\phi(t_{2},x_{2})\rangle\\
\nonumber
&=-\frac{1}{4\pi}\left\{\ln[\ri\mu(t_{12}-x_{12})+\epsilon]-\ln[\ri\mu(t_{12}+x_{12})+\epsilon]\right\}
\end{align}
where $t_{12}=t_1-t_2$, $x_{12}=x_{1}-x_{2}$, and $\mu>0$ is an infrared cutoff that does not affect physical (anyonic) correlators that are invariant under the shift symmetries (\ref{shift}), provided that we properly choose  the normalization
\begin{equation}
\nonumber
z=\mu^{(\zeta_{+}^2+\zeta_{-}^2)/4}\,.
\end{equation}
 This is important since (\ref{shift}) are just the bosonic version (up to an overall normalization)
of the $U_{V,A}(1)$ chiral symmetries (\ref{vectorial}, \ref{axial}).
For instance, we have
\begin{align}
\label{corrPsi1}
& \langle\psi_{1}^{*}(t_{1},x_{1})\psi_{1}(t_{2},x_{2})\rangle= \langle\psi_{1}(t_{1},x_{1})\psi^{*}_{1}(t_{2},x_{2})\rangle=  \frac{1}{2\pi}\left[\mathcal{D}(vt_{12}-x_{12})\right]^{\frac{(\zeta_{+}-\zeta_{-})^{2}}{4}} 
\left[\mathcal{D}(vt_{12}+x_{12})\right]^{\frac{(\zeta_{+}+\zeta_{-})^{2}}{4}}\\
\label{corrPsi2}
& \langle\psi_{2}^{*}(t_{1},x_{1})\psi_{2}(t_{2},x_{2})\rangle=\langle\psi_{2}(t_{1},x_{1})\psi^{*}_{2}(t_{2},x_{2})\rangle=  \frac{1}{2\pi}\left[\mathcal{D}(vt_{12}-x_{12})\right]^{\frac{(\zeta_{+}+\zeta_{-})^{2}}{4}} 
\left[\mathcal{D}(vt_{12}+x_{12})\right]^{\frac{(\zeta_{+}-\zeta_{-})^{2}}{4}}
\end{align}
where $\mathcal{D}(x)=(\ri x+\epsilon)^{-1}$ and $t_{12}=t_1-t_2$, $x_{12}=x_{1}-x_{2}$. 
The other 2-point functions vanish by $U(1)_{V}\times U(1)_{A}$ symmetry \footnote{We select the physical Hilbert space by means of the neutrality conditions, $\sum {\zeta_{\pm}}=0$ on the n-point correlation functions, such that $U(1)_{V}$ and $U(1)_{A}$ hold.}.
Also, we see that correlation functions are invariant under dilatations
\begin{equation}
\label{scale}
\psi_{\alpha}(t,x)\rightarrow \lambda^{(\zeta_{+}^2+\zeta_{-}^{2})/4}\psi_{\alpha}(\lambda t, \lambda x)\qquad \lambda>0
\end{equation}
and Lorentz boost $\Lambda_{\chi}$ that leaves the sound-cone $v^2t^2-x^2$ invariant\footnote{The Lorentz transformation that leaves the sound cone invariant is defined as in (\ref{boost00}) where $v_{F}$ is replaced everywhere by $v$. For the moment $v$ is just a free parameter that will be fixed in the next sections in terms of the coupling constants.}
\begin{equation}
\label{boost}
\psi_{\alpha}(t,x)\rightarrow e^{-(-1)^{\alpha}\chi \k/2}\psi_{\alpha}( \Lambda_{\chi}(t,x))
\end{equation}
where $\chi$ is the associated rapidity.
 From (\ref{scale}, \ref{boost}) we get respectively the dimension and the spin
\begin{equation}
\label{dimension-spin}
[\psi]=(\zeta_{+}^2+\zeta_{-}^{2})/4\,,\quad  s(\psi)=\pm \k/2\,.
\end{equation}
Free anyons have $\zeta_{+}=-\zeta_{-}$ and $[\psi]=\k/2$.
Canonical free fermions with spin $\pm 1/2$ and dimension $1/2$ correspond to $\zeta_{+}=-\zeta_{-}=\pm 1$.

We stress that the perturbation of free anyon dynamics by a mass term $m\bar{\psi}\psi$, which is invariant under the Lorentz boost (\ref{boost}), does not correspond to a system described by the massive Dirac equations
\begin{equation}
\label{eom2}
\ri\left(\partial_t+\partial_{x}\right)\psi_{1}=m\psi_{2}\,,\quad \ri\left(\partial_t-\partial_{x}\right)\psi_{2}=m\psi_{1}\,.
\end{equation} 
Indeed, the right- and left-hand sides of these equations transform in different ways under the Lorentz boosts (\ref{boost}). In practice, there is a mismatch (unless $\psi$'s are canonical fermions) between the spin counting on the two sides of these equations: $\mp(1-\k/2)$ on the left and $\mp\k/2$ on the right. 
With a slight abuse of language, we will keep referring to the deformation by $(\bar{\psi}\psi)$ as  a \textit{mass term} perturbation.

\subsection{Composite operators}

We are  now  ready to express the anyonic deformations (\ref{Odeform}) in the bosonized language. In particular, we are interested in conserved currents and operators that preserve $\mathcal{C}$, $\mathcal{P}$ and $\mathcal{T}$.

\subsubsection{Charges and currents}
Vectorial and axial $U(1)$ transformations (\ref{vectorial}, \ref{axial}) define
 two current densities, $J_{\mu}=(\rho_{+}, j_{+})$ and $J^5_{\mu}=(\rho_{-},j_{-})$, in terms of 2-anyon composite operators, $\bar{\psi}\gamma_{\mu}\psi$ and $\bar{\psi}\gamma_{\mu}\gamma^5\psi$ respectively. 
 After removing the UV divergences coming from the product of coincident anyons, we are left with linear derivatives of the bosonic fields. For instance, the current associated to $U(1)_{V}$ is given by
 \begin{align}
 \rho_{+}(t,x) & =\frac{1}{\sqrt{\pi}\zeta_{+}}(\partial_{x}\tilde{\phi})(vt,x)\\
 j_{+}(t,x) & =\frac{v}{\sqrt{\pi}\zeta_{+}}(\partial_{t}\tilde{\phi})(vt,x)
 \end{align}
 where the overall finite normalization is fixed by Ward identities (\ref{WardV}, \ref{WardA}), 
 and the current conservation equation
\begin{equation}
\partial_{t}\rho_{+}-\partial_{x}j_{+}=0\,.
\end{equation}
Similarly, we get the axial currents
 \begin{align}
 \rho_{-}= & -\frac{1}{\sqrt{\pi}\zeta_{-}}(\partial_{x}\phi)(vt,x)\\
 j_{-}= & -\frac{v}{\sqrt{\pi}\zeta_{-}}(\partial_{t}\phi)(vt,x)\,.
 \end{align}

We can write these currents in a Lorentz covariant way, namely
\begin{align}
\label{Hodgecurrents}
J_{\mu}=-\frac{1}{\sqrt{\pi}\zeta_{+}}\epsilon_{\mu\nu}\partial^\nu\,\tilde{\phi}(vt,x)\,,\quad J^5_{\mu}=\frac{1}{\sqrt{\pi}\zeta_{-}}\epsilon_{\mu\nu}\,\partial^\nu\phi(vt,x)\,.
\end{align} 
 Of course there is no anomalous dimension generated for conserved currents, $[J_{\mu}]=[J^5_{\mu}]=1$.
 Note that the classical relations $j_{-}=-\rho_{+}$ and $j_{+}=-\rho_{-}$ are broken at the quantum level by renormalization effects which replace them with
 \begin{align}
 \label{qrela}
 \rho_{-}=\frac{\zeta_{+}}{\zeta_{-}v}j_{+}\,,\quad \rho_{+}=\frac{\zeta_{-}}{\zeta_{+}v}  j_{-}\,.
 \end{align}

\subsubsection{Mass term and 4-anyon operators}

 The mass term $\bar{\psi}\psi=\psi_{1}^*\psi_2+\psi^*_2\psi_1$ is the only relevant 2-anyon operator that preserves $\mathcal{C}$, $\mathcal{P}$, and $\mathcal{T}$ (see App.\ref{deform}). It also mixes the chiralities breaking $U(1)_{A}$. 
 In terms of bosonic fields this term is given by
\begin{align}
\label{cos}
\bar{\psi}\psi=\mu^{\zeta_{-}^2}:\cos\left[2\sqrt{\pi}\zeta_{-}(\tilde{\phi}-Q)+\frac{\pi}{2}\k\right]:\,.
\end{align}

The product of two conserved currents at the same point has very simple UV behavior because it involves the product of two free bosonic fields. Thus, all 4-anyon operators of the form $\rho_{\pm}\rho_{\pm}$, $\rho_{\pm}j_{\pm}$ and $j_{\pm}j_{\pm}$, are well defined once we take the normal ordering, $\sim:\partial\phi\partial\phi:$. This turns out to be the reason why Tomonaga-Luttinger and Thirring models are exactly solvable.

The short distance behavior of $(\bar{\psi}\psi)^2$ is also quite simple. Indeed, from (\ref{cos}) we get \footnote{This is just the quantum version of the trigonometrical identities $\cos^2x=\frac{1}{2}\cos(2x)+1/2$, $\sin(2x)=2\sin x\cos x$  and $\cos^2 x+\sin^2 x=1$, where the constant terms are now UV divergent c-numbers that renormalize the vacuum energy without affecting the dynamics.}
\begin{align}
\label{4anyons}
(\bar{\psi}\psi)^2=\mu^{4\zeta_{-}^2}:\cos\left[4\sqrt{\pi}\zeta_{-}(\tilde{\phi}-Q)+\pi\k\right]:\,\,.
\end{align}
This result implies that $(\bar{\psi}\psi)$- and $(\bar{\psi}\psi)^2$-insertions can be treated 
on the same footing by rescaling the $\beta$ parameter in the sine-Gordon model.

\section{Dualities}
\label{sect-duality}

In this section we discuss the dualities between the Tomonaga-Luttinger, Thirring and Schwinger models for anyons (and their renormalizable deformations) with respect to the sine-Gordon model.

\subsection{Tomonaga-Luttinger model}
\label{Tomonaga-Luttinger}

The anyonic Tomonaga-Luttinger model is defined in terms of charge-charge interactions\footnote{Often the couplings $g_{2,4}=2\pi(g_{+}\mp g_{-})$ are used in the literature.} 
\begin{equation}
\label{TL}
\mathcal{L}_{TL}=-\pi g_{+}\rho_{+}^2-\pi g_{-}\rho_{-}^2\,.
\end{equation} 
In order to avoid extra divergences it is convenient to look directly at the equations of motion
\begin{align}
\ri (\partial_t +\partial_x) \psi_1(t,x) = &  2\pi :\left[g_+\, \rho_+(t,x) + g_-\, \rho_-(t,x) \right]\psi_1(t,x) : \\
\ri (\partial_t -\partial_x) \psi_2(t,x) = &  2\pi :\left[g_+\, \rho_+(t,x) - g_-\, \rho_-(t,x) \right]\psi_2(t,x) : 
\label{eqm1}
\end{align}
They are solved using bosonization by the following choices \cite{Bellazzini:2008fu}
\begin{align}
\zeta_{+}^2= & |\k|\sqrt{\frac{\k +2g_{+}}{\k +2g_{-}}}\\
\zeta_{-}^2= & |\k|\sqrt{\frac{\k +2g_{-}}{\k +2g_{+}}}
\label{zetamTL} \\
v= & \sqrt{(1 +2g_{-}/\k)(1+2g_{+}/\k)}\,.
\label{vTL}
\end{align}
As anticipated, $\zeta_{\pm}$ and $v$ now depend  on the interactions and the statistical parameter, generalizing the well known expression for canonical fermions in Tomonaga-Luttinger model. The traditionally used parameter $K$ \cite{Giamarchi} in our notation coincides at $\k=1$ with $\zeta_{-}^2$.
While anyonic statistics $\k\neq\pm 1$ are not directly visible on the speed of the excitations (by 
rescaling the couplings we can absorb the $\k$-dependence), their impact is visible on the correlation functions that depend on $\zeta_{\pm}$. For instance,  the 2-point functions
\begin{align}
\nonumber
W_{\alpha\alpha}(t_{12},x_{12},\k, g_{+}, g_{-})=\langle\psi_{\alpha}^*(t_1,x_{1})\psi_{\alpha}(t_{2},x_{2})\rangle\,,
\end{align}
for generic anyonic statistics and couplings have a simple scaling property in $\k$
\begin{equation*}
W_{\alpha\alpha}(t_{12},x_{12},,\k, g_{+}, g_{-})=\left[W_{\alpha\alpha}(t_{12},x_{12},1, \frac{g_{+}}{\k}, \frac{g_{-}}{\k})\right]^{|\k|}\,.
\end{equation*}
This  equation relates the  2-point functions of the canonical fermionic Tomonaga-Luttinger model with their anyonic analog.

\subsubsection{Deformations and the sine-Gordon model}

Let us now add one of the possible deformations given in  (\ref{Odeform}). The terms with the current-current interactions just give rescaling of $g_{\pm}$. Both $\bar{\psi}\psi$ and $(\bar{\psi}\psi)^2$ terms correspond in perturbation theory to the insertion of  cosine terms.  For instance, the mass term  gives $\cos[2\sqrt{\pi}\zeta_{-}(\tilde{\phi}-Q)+\frac{\pi}{2}\k]$. Since we are in fact perturbing a theory expressed in terms of free massless bosons, we can always shift $\tilde{\phi}$ such that $Q$ and $\k$ disappear from the argument of the cosine. Then, our massive deformation of the Tomonaga-Luttinger is equivalent (up to matching of the renormalization scale) to the sine-Gordon model with the  Hamiltonian density
\begin{align}
\mathcal{H}_{sG}=&\frac{v}{2}:\left[\Pi^2+(\partial_{x}\tilde{\phi})^2\right]:-\frac{m^2}{\beta^2}:\cos(\beta\tilde{\phi}):\\
\beta^2= & 4\pi\zeta_{-}^2=4\pi |\k|\left(\frac{\k +2g_{-}}{\k +2g_{+}}\right)^{1/2}\,.
\label{betaTL}
\end{align}
where $\Pi(t,x)=-\partial_{x}\phi(vt,x)$ is the conjugate momentum of $\tilde{\phi}(vt,x)$, see (\ref{nonlocPh}).
The same arguments  apply for 4-anyon operators.

From the seminal work of Coleman \cite{Coleman:1974bu} we know that $\beta^2 < 8\pi$ in order for the sine-Gordon model to have a stable vacuum i.e. the energy spectrum bounded from below.  This constraint simply states that the dimension $\zeta_{-}^2$ of $\bar{\psi}\psi$ has to be less than $2$. Putting this together with the reality of $\zeta_{\pm}$, we set the non trivial range where the vacuum is stable
\begin{equation}
0<\left(\frac{\k +2g_{-}}{\k +2g_{+}}\right)<\frac{4}{\k^2}\,.
\label{thetabound}
\end{equation}
One can look at (\ref{thetabound}) as a constraint on the coupling constants for fixed statistics: it says that
one coupling has to dominate over the other for an amount fixed by $\k$.

\subsection{Thirring model}
\label{sect-Thirring}

The Thirring model \cite{Thirring:1958in} describes a Lorentz invariant 4-anyon  interaction and is defined by 
\begin{equation}
\mathcal{L}_{Th}=-\pi g J^{\mu} J_{\mu}=-\pi g \left( \rho_{+}^2-j_{+}^{2}\right)\,.
\end{equation}
Again, in order to avoid extra divergences, it's useful to look at the equations of motion
\begin{align}
\ri (\partial_t +\partial_x) \psi_1 =& 2\pi g :\left( \rho_+ +j_{+} \right)\psi_1 : \,,\\
\ri (\partial_t -\partial_x) \psi_2 =& 2\pi g:\left(\rho_+ - j_{+}\right)\psi_2 : \,.
\label{eqm2}
\end{align}
 Recalling the relations (\ref{qrela}) between charges and currents, we see that the Thirring model (and hence Lorentz symmetry) is recovered from the Tomonaga-Luttinger model by tuning the coupling constants to $g_{+}=g$ and $g_{-}=g v \zeta_{-}/\zeta_{+}$. In particular, we find that the model is solved by these choices \cite{lm-99}
\begin{align}
\label{Th1}
\zeta_{-}= & \pm \frac{\k}{\sqrt{\k +2g}}\,,\\
\label{Th2}
\zeta_{+}= & \mp\sqrt{\k+2g}\,,\\
v= & 1\,.
\label{vTH}
\end{align}
Non surprisingly, the excitations travel at the Fermi velocity $v=v_{F}=1$ since Lorentz symmetry is respected.
From the analysis of the Tomonaga-Luttinger model we conclude that the massive Thirring model is perturbatively equivalent to a sine-Gordon model with a $\cos\beta\tilde{\phi}$ potential where 
\begin{equation}
\beta^2=4\pi \zeta_{-}^2=\frac{4\pi\k^2}{\k+2g}\,.
\label{betaTh}
\end{equation}
The vacuum stability of the sine-Gordon model is now given by
\begin{equation}
g> \k(\k-2)/4
\label{stabilityTh}
\end{equation}
that is a stronger bound than the reality condition $g>-\k/2$ for $\zeta_{\pm}$.
For canonical fermions, $\k=1$, we recover the Coleman's bound.

\subsection{Schwinger model}

The Schwinger model describes the quantum electrodynamics of a charged particle in 1D,
\begin{equation}
\mathcal{L}_{Sc}=-\frac{1}{4}F_{\mu\nu}F^{\mu\nu}-e A_{\mu}J^{\mu}\,.
\end{equation}
Choosing the gauge $A_x=0$ it is clear that there is no physical degree of freedom propagating for the photon.
Indeed, the equation of motion for $A_t$ just sets a constraint,
\begin{equation}
\label{constr1}
\partial_{x}^2 A_t=-e\rho_{+}\,.
\end{equation}
Integrating by parts and using the constraint (\ref{constr1}), we get
 \begin{equation}
\mathcal{L}_{Sc}\rightarrow -\frac{1}{2}(\partial_{x}A_{t})^2\,.
\end{equation}
Bosonization yields
 $\rho_{+}=1/(\sqrt{\pi}\zeta_{+})\partial_{x}\tilde{\phi}$, so
\begin{equation}
\partial_{x}A_{t}=-\frac{e}{\sqrt{\pi}\zeta_{+}}\tilde{\phi}-\mathcal{E}\,,
\end{equation}
where $\mathcal{E}$ is an integration constant. Thus, after shifting $\tilde{\phi}$ by an amount $-\sqrt{\pi}\zeta_{+}\mathcal{E}/e$,  the system is dual to a free massive Klein-Gordon model
\begin{equation}
\mathcal{H}=\frac{v}{2}:\left[\Pi^2+(\partial_{x}\tilde{\phi})^2\right]:+\left(\frac{e^2}{2\pi\zeta^2_{+}}\right):\tilde{\phi}^2:
\end{equation}
with mass 
\begin{equation}
\label{massSchwinger}
m^2=\frac{e^2}{\pi\zeta^2_{+}v^{3}}\,.
\end{equation}
We recover the standard Schwinger result \cite{Schwinger:1962tp} for a relativistic canonical fermion by setting $\zeta_{+}^2=1$ and $v=v_{F}=1$.
The presence of this \textit{meson} confirms the confining nature of $U(1)$ in 1D where the potential $V$  between two localized charges grows linearly
\begin{equation}
V=\langle y|\frac{1}{\partial_{x}^{2}}|x\rangle=\frac{1}{2}|x-y|\,.
\end{equation}

\subsubsection{Deformations}

Let us now add a mass deformation, $\bar{\psi}\psi$. In the bosonic dual theory, it
 corresponds  to a term $\sim\cos[2\sqrt{\pi}\zeta_{-}(\tilde{\phi}-Q)+\frac{\pi}{2}\k]$.  
 After a shift in $\tilde\phi$, the system is equivalent to a massive sine-Gordon model  with
\begin{equation}
\cos\left[2\sqrt{\pi}\zeta_{-}(\tilde\phi-Q)+\frac{\pi}{2}\tilde{\k}\right]
\end{equation}
where $\tilde{\k}=\k(1-4\mathcal{E}/e)$.

\subsubsection{The chiral anomaly}

We still get solvable models if we add Thirring or Tomonaga-Luttinger interactions to the Schwinger lagrangian.
For simplicity we focus on the Schwinger-Thirring model where $v=1$ by Lorentz symmetry. By means of bosonization and (\ref{Th1}-\ref{vTH}) we get the correlation functions of the theory and the mass of the bound state reads
\begin{equation}
\label{massSchTH}
m^2=\frac{e^2}{\pi(\k+2g)}\,.
\end{equation}

Another interesting property of the anyonic Schwinger-Thirring model is the modification of the chiral anomaly.
It is best to do this calculation in Lorentz gauge, $\partial_{\mu} A^{\mu}=0$.
Defining the vectorial and axial currents $J_{\mu}$ and $J_{\mu}^5$ by point-splitting, we enforce the $U(1)_{V}$ gauge symmetry by inserting Wilson lines in the definitions (\ref{psi1}, \ref{psi2}). Thus, we end up with the following expressions
\begin{align}
J_{\mu}=& -\frac{1}{\sqrt{\pi}\zeta_{+}}\partial_{\mu}\phi-\frac{e}{\pi\zeta_{+}^2}A_{\mu}\\
J^5_{\mu}=&\frac{\zeta_{+}}{\zeta_{-}}\epsilon_{\mu\nu}J^{\nu}=-\frac{1}{\sqrt{\pi}\zeta_{-}}\epsilon_{\mu\nu}\partial^{\nu}\phi+\frac{e}{\pi\k}\epsilon_{\mu\nu}A^{\nu}
\end{align}
and the chiral anomaly simply reads
\begin{equation}
\label{chiralanomaly}
\partial^{\mu} J^5_{\mu}=\frac{e}{2\pi\k}\epsilon_{\mu\nu}F^{\mu\nu}\,.
\end{equation}
It is important to stress that even non-canonical fermions (i.e. with odd statistical parameter but $\k\neq 1)$, 
modify chiral anomaly  (\ref{chiralanomaly}) from the standard Schwinger result at $\k=1$. Conversely, there is no contribution from the other non-electromagnetic interactions \footnote{Here we disagree with the earlier result $\partial^{\mu} J^5_{\mu}=\frac{e}{2\pi(1+2 g)}\epsilon_{\mu\nu}F^{\mu\nu}$ of Ref. \cite{Georgi:1971iu} 
 for canonical fermions ($\k=1$). The difference with our (\ref{chiralanomaly}) comes from the normalization of $J^5_\mu$. They defined $J^5_{\mu}=\epsilon_{\mu\nu}J^{\nu}$ while we are fixing the normalization demanding the Ward identity (\ref{WardA}) in the limit of vanishing electric charge $e\rightarrow0$. These two definitions are equivalent only at the classical level, see Eq.(\ref{Hodgecurrents}, \ref{qrela}).}.

The equations of motion for $A_{\mu}$ now imply
\begin{equation}
(\square+\frac{e^2}{\pi\zeta_{+}^2})\epsilon^{\mu\nu}F_{\mu\nu}=0\,,
\end{equation}
confirming again the presence of a composite state of mass (\ref{massSchwinger}, \ref{massSchTH}).

\section{Conclusions}
\label{sect-conclusions}

We have discussed the low-energy dynamics of anyons in (1+1)-dimensions with the smallest number of derivatives and the most general renormalizable interactions that are $\mathcal{C}$, $\mathcal{P}$, and $\mathcal{T}$ symmetric. 
Despite the presence of Lorentz violating interactions (as in the Tomonaga-Luttinger model), Lorentz symmetry with respect to the sound-cone is always recovered at low energy. Furthermore, the most general  anyonic interactions are dual to the sine-Gordon model for bosons, with coupling constants depending on the statistical parameter (\ref{betaTL},\ref{betaTh}). The stability of the vacuum itself depends on the statistics (\ref{thetabound},\ref{stabilityTh}).
We also discussed the anyonic realization of the Schwinger model where the mass of the composite state (\ref{massSchwinger}, \ref{massSchTH}) and the chiral anomaly (\ref{chiralanomaly}) are corrected by the statistical parameter.

\section*{Acknowledgment}
I'd like to thank Andr\'e Leclair, Mihail Mintchev, Alvise Varagnolo and Flip Tanedo for reading and commenting on the paper. I'm also grateful to Henry Tye and Csaba Cs\'aki for useful discussions. The research of the author has been supported in part by the NSF grant PHY-0757868.

\appendix

\section{Deformations}
\label{deform}

In this appendix we give the full classification of canonically renormalizable deformations according to their transformation properties under $\mathcal{C}$, $\mathcal{P}$, and $\mathcal{T}$
\begin{align}
\nonumber
\mathcal{P}\psi_{1}(t,x)\mathcal{P}^{-1}=&\psi_{2}(t,-x)& \,\,\, \mathcal{P}\psi_{2}(t,x)\mathcal{P}^{-1}=&\psi_{1}(t,-x)&\\
\nonumber
\mathcal{C}\psi_{1}(t,x)\mathcal{C}^{-1}=&\psi_{1}^*(t,x) &\,\,\,  \mathcal{C}\psi_{2}(t,x)\mathcal{C}^{-1}=&-\psi_{2}^*(t,x)& \\
\nonumber
\mathcal{T}\psi_{1}(t,x)\mathcal{T}^{-1}=&\psi_{2}(-t,x) &\,\,\,  \mathcal{T}\psi_{2}(t,x)\mathcal{T}^{-1}=&\psi_{1}(-t,x)&
\end{align}
where $\mathcal{T}$ it is an antiunitary transformation. We focus only on deformations that do not break electromagnetism i.e. preserve $U(1)_{V}$. 

From (\ref{dimension-spin}) we can read the dimension of free anyons, $[\psi]=|\k|/2$. Then the canonical dimensions of 
\begin{equation}
\label{deformcounting}
(\psi^{*}_{\alpha}\psi_{\beta})\,,\,\,(\psi^{*}_{\alpha}\psi_{\beta}\psi^{*}_{\gamma}\psi_{\delta})\,,\,\, (\psi^{*}_{\alpha}\partial_{\mu}\psi_{\beta})
\end{equation} 
are respectively $|\k|$, $2|\k|$, and $|\k|+1$.
Those operators are canonically renormalizable when $|\k|\leq 1$. Actually, no other operator is allowed in the range $2/3<|\k|\leq 1$. We restrict our classification to this case just to deal with a  finite number of possible perturbations.  Of course, canonical dimensions may differ from the actual dynamical dimensions. For instance, conserved currents always have dimension 1 so that the deformations as in the Tomonaga-Luttinger, Thirring, and Schwinger  models are allowed for any real $\k$. Moreover, the Tomonaga-Luttinger and the Thirring models define conformal field theories with arbitrarily large couplings so that the dynamical dimensions may get big corrections from the canonical ones. For instance, $[\bar{\psi}\psi]=\zeta^2_{-}$ and $[(\bar{\psi}\psi)^2]=4\zeta^2_{-}$ with $\zeta_{-}$ given in (\ref{zetamTL}) and (\ref{Th1}). While the present classification is valid in the small coupling regime, other classifications around those fixed points with large couplings are possible along the same lines.

\subsection{Relevant and marginal deformations}

There are four Hermitian deformations $\mathcal{O}$ with canonical dimension $[\mathcal{O}]=|\k|$, built out of linear combinations of the anyon bilinears $\psi^{*}_{\alpha}\psi_{\beta}$. 
In order to avoid the ambiguities about the ordering of the anyon fields at the same point in $\mathcal{O}$, we check the symmetry content by looking at the  equations of motion associated with the modified lagrangian. 
Thus,  $\mathcal{O}_{m_{+}}$ and $\mathcal{O}_{m_{-}}$ 
\begin{align}
\nonumber
\mathcal{O}_{m_{+}}= \psi^*_{1}\psi_{2}+\psi^*_{2}\psi_{1}\,,\qquad
\mathcal{O}_{m_{-}}= \ri\left(\psi^*_{1}\psi_{2}-\psi^*_{2}\psi_{1}\right)\,,
\end{align}
preserve $\mathcal{C}\mathcal{P}\mathcal{T}$ while $\rho_{+}$ and $\rho_{-}$,
\begin{align}
\nonumber
\rho_{+}= \psi^*_{1}\psi_{1}+\psi^*_{2}\psi_{2}\,,\qquad
\rho_{-}= \psi^*_{1}\psi_{1}-\psi^*_{2}\psi_{2}
\end{align}
break it explicitly. 
Note also that demanding just $\mathcal{C}\mathcal{P}\mathcal{T}$-symmetric deformations, 
we recover Lorentz symmetry as an accidental symmetry of the action. This is immediately visible writing $\mathcal{O}_{m_{\pm}}$ as Lorentz scalars, $\mathcal{O}_{m_+}=\bar{\psi}\psi$ and $\mathcal{O}_{m_-}=-\ri\bar{\psi}\gamma^5\psi$.
It is also clear that $\rho_{+}=\bar\Psi\gamma^0\Psi$ and $\rho_{-}=\bar{\Psi}\gamma^1\Psi$ break Lorentz symmetry in the lagrangian $\delta\mathcal{L}=c_{+}\rho_{+}+c_{-}\rho_{-}=A_{\mu}\bar{\Psi}\gamma^\mu\Psi$, because of the external vectorial field $A_{\mu}=(c_{+},c_{-})$.
The following table summarizes the symmetry properties of these 2-anyon deformations
\begin{equation}
\nonumber
\begin{array}{l|c|c|c|c|c|c}
\mbox{Relevant operators} & \mathcal{C} & \mathcal{P}  &  \mathcal{T} & U(1)_{A} & L &\mathcal{C}\mathcal{P}\mathcal{T}\\
\hline
\mathcal{O}_{m_+}=\psi^*_{1}\psi_{2}+\psi^*_{2}\psi_{1} & + & + & + & - & + & +\\
\mathcal{O}_{m_-}=\ri\left(\psi^*_{1}\psi_{2}-\psi^*_{2}\psi_{1}\right) & - & - & + & - & + & +\\
\hline
\rho_{+}=\psi^*_{1}\psi_{1}+\psi^*_{2}\psi_{2} & - & + & + & + & - & -\\
\rho_{-}=\psi^*_{1}\psi_{1}-\psi^*_{2}\psi_{2} & - & - & - & + & - & -
\end{array}
\end{equation}

There are six possible (Hermitian) operators with $[\mathcal{O}]=|\k|+1$ that contain derivatives. 
 Two of them are just a trivial rescaling of the Fermi velocity $v_{F}$
that can be absorbed by changing the time or spatial scale. 
Thus, we are left with four of those deformations that we organize in two 
subsets, $D_{i}$ and $E_{i}$,  that preserve or break $\mathcal{C}\mathcal{P}\mathcal{T}$ respectively. 
Their explicit expressions are given in the following table:
\begin{equation}
\nonumber
\begin{array}{l|c|c|c|c|c|c}
\mbox{Marginal with deriv.} & \mathcal{C} & \mathcal{P}  &  \mathcal{T} & U(1)_{A} & L &\mathcal{C}\mathcal{P}\mathcal{T}\\
\hline
D_{1}=\psi^*_1\ri\partial_{t}\psi_{1}-\psi^*_2\ri\partial_{t}\psi_{2} & + & - & - & + & - & +\\
D_{2}=\psi^*_1\ri\partial_{x}\psi_{1}+\psi^*_2\ri\partial_{x}\psi_{2} & + & - & - & + & - & +\\
\hline
E_{1}=\psi^*_1\ri\partial_{t}\psi_{2}+\psi^*_2\ri\partial_{t}\psi_{1} & - & + & + & - & - & -\\
E_{2}=\psi^*_1\ri\partial_{x}\psi_{2}+\psi^*_2\ri\partial_{x}\psi_{1} & - & - & - & - & - & -
\end{array}
\end{equation}
The main difference with respect to the 2-anyon deformations without derivatives is that enforcing 
$\mathcal{C}\mathcal{P}\mathcal{T}$ does not guarantee the emergence of Lorentz symmetry at low energy. 
In order to recover this, we need to impose a slightly stronger symmetry as $\mathcal{C}\mathcal{P}$ and $\mathcal{T}$, or 
$\mathcal{C}\mathcal{T}$ and $\mathcal{P}$. 

Let us now consider 4-anyon operators with no derivatives.  There are ten such Hermitian $4$-anyon interactions,  
as shown in  the following table: 
\begin{align}
\nonumber
\begin{array}{l|c|c|c|c|c|c}
\mbox{Marginal without deriv.} & \mathcal{C} & \mathcal{P}  &  \mathcal{T} & U(1)_{A} & L &\mathcal{C}\mathcal{P}\mathcal{T}\\
\hline
\mathcal{O}_{m_+}\mathcal{O}_{m_+}=(\bar{\psi}\psi)^2  & + & + & + & - & + & +\\
\mathcal{O}_{m_-}\mathcal{O}_{m_-}=-(\bar{\psi}\gamma^{5}\psi)^2  & + & + & + & - & + & +\\
\mathcal{O}_{m_+}\mathcal{O}_{m_-}=-\ri(\bar{\psi}\psi)(\bar{\psi}\gamma^{5}\psi)  & - & - & + & - & + & +\\ 
J^{\mu}J_{\mu}=(\bar{\psi}\gamma^{\mu}\psi)(\bar{\psi}\gamma_{\mu}\psi)& + & + & + & + & + & +\\
\hline
TL=g_{+}\rho_{+}^2+g_{-}\rho_{-}^2  & + & + & + & + & - & +\\
\rho_{+}\rho_{-} & +  & - & - & + & - & +\\
\hline
\mathcal{O}_{m_+}\rho_{+}  & - & + & + & - & - & -\\
\mathcal{O}_{m_+}\rho_{-}  & - & - & - & - & - & -\\
\mathcal{O}_{m_-}\rho_{+}  & +  & - & + & - & - & -\\
\mathcal{O}_{m_-}\rho_{-}  & + & + & - & - & - & -\\
\end{array}
\end{align}
Six of these operators respect $\mathcal{C}\mathcal{P}\mathcal{T}$.  
Imposing a stronger discrete symmetry as $\mathcal{C}\mathcal{P}$ and $\mathcal{T}$, 
we can remove all Lorentz-violating deformations but the Tomonaga-Luttinger interactions
$(g_{+}\rho_{\pm}^2+g_{-}\rho_{-}^2)$. However, Tomonaga-Luttinger interactions break the Lorentz symmetry in a very special way, preserving in fact a Lorentz symmetry for the sound-cone $v^2 t^2-x^2$. 
Note that 4-anyon operators built out of product of conserved currents give rise to exactly marginal deformations for any $\k$.

We end this appendix recalling that $(\bar{\psi}\gamma^5\psi)^2$ can be obtained by means of a linear combination of  $(\bar{\psi}\gamma^\mu\psi)(\bar{\psi}\gamma_\mu\psi)$ and $(\bar{\psi}\psi)^2$. Hence, the operators $\bar{\psi}\psi$, $(\bar{\psi}{\psi})^{2}$, $(\bar{\psi}\gamma^\mu\psi)(\bar{\psi}\gamma_\mu\psi)$, and $\rho_{\pm}^2$
form a basis that respects $\mathcal{C}$, $\mathcal{P}$ and $\mathcal{T}$.

\section{The $\k$-commutator}
\label{app-kcomm}

We discuss now the following $\k$-commutator at equal times
\begin{equation}
\label{thetacommutator}
[\psi^*(t,x_1),\psi(t,x_2)]_{\k}\equiv \psi^*(t,x_1)\psi(t,x_2)-e^{\ri\pi\k\varepsilon(x_1-x_2)}\psi(t,x_2)\psi^*(t,x_1)\,.
\end{equation}
Let us focus for simplicity on $\psi=\psi_1$, analogous results hold for $\psi_2$ as well.
Using bosonized expression (\ref{psi1}) for $\psi_1$ together with the commutation relation (\ref{nonlocPh}) between $\phi$ and $\tilde{\phi}$, it is easy to show that (\ref{thetacommutator}) vanishes for $x_1\neq x_2$.  
However, at coincident points, the $\k$-commutator between $\psi^*$ and $\psi$ becomes singular and some care is needed.
In particular it is convenient to isolate the source of singularity coming from the vacuum expectation value
\begin{equation}
\label{thetacommExp}
[\psi^*(t,x_1),\psi(t,x_2)]_{\k}=\langle [\psi^*(t,x_1),\psi(t,x_2)]_{\k}\rangle 
:e^{-\ri\sqrt{\pi}[\zeta_{+}(\phi(x_1)-\phi(x_2))-\zeta_{-}(\tilde{\phi}(x_1)-\tilde{\phi}(x_2))] }:\,.
\end{equation}
whereas the exponential term under the normal ordering is regular for $x_1\rightarrow x_{2}$.
From the correlation functions (\ref{corrPsi1}) we get
\begin{align}
\nonumber
\langle [\psi^*(t,x_1),\psi(t,x_2)]_{\k}\rangle= & \frac{1}{2\pi}|x_{1}-x_{2}|^{-\frac{(\zeta_{+}+\zeta_{-})^2}{2}}
\left[ \frac{1}{\left(-\ri(x_{1}-x_{2})+\epsilon\right)^\k}- \frac{1}{\left(-\ri(x_{1}-x_{2})-\epsilon\right)^\k}\right]\\
\label{thetacommVEV}
= & \frac{\ri}{\pi}\sin(\frac{\pi\k}{2})\varepsilon(x_{1}-x_{2})|x_{1}-x_{2}|^{-2[\psi]}\qquad [\psi]=(\zeta_{+}^2+\zeta_{-}^2)/4
\end{align}
where $[\psi]$ is the scaling dimension of $\psi$ (see Eq.(\ref{dimension-spin})). 
In the limit of half-integer dimension, (\ref{thetacommVEV}) simplifies to a certain derivative of the delta function \cite{Gelfand}
\begin{equation}
\langle [\psi^*(t,x_1),\psi(t,x_2)]_{\k}\rangle=-\ri\sin(\frac{\pi\k}{2})\frac{1}{(2[\psi]-1)!}\delta^{(2[\psi]-1)}(x_{1}-x_{2})\qquad 
2[\psi]\in\NN^+\,.
\end{equation} 
Now, we expand the exponential in (\ref{thetacommExp})  in powers of $(x_{1}-x_{2})$ using (\ref{Hodgecurrents})
\begin{equation}
:e^{-\ri\sqrt{\pi}[\zeta_{+}(\phi(x_1)-\phi(x_2))-\zeta_{-}(\tilde{\phi}(x_1)-\tilde{\phi}(x_2))] }:=
1-\ri\pi\k(x_{1}-x_{2})(\rho_{+}(x_2)+\rho_{-}(x_2))+\ldots
\end{equation}
and we put it together with (\ref{thetacommVEV})
\begin{equation}
\label{OPEtheta}
[\psi^*(t,x_1),\psi(t,x_2)]_{\k}=
\sin(\frac{\pi\k}{2})\left(\frac{\ri}{\pi}\frac{\varepsilon(x_{1}-x_{2})}{|x_{1}-x_{2}|^{2[\psi]}}+
\frac{\rho_{+}(x_2)+\rho_{-}(x_2)}{|x_{1}-x_{2}|^{2[\psi]-1}}+\ldots\right)
\end{equation}
This expression is nothing but the explicit operator product expansion of the $\k$-commutator where ellipses contain a finite number of local operators $\mathcal{O}_{n}(x_{2})$ divided by positive powers of $(x_{1}-x_{2})$ decreasing by integer steps. The operators $\rho_{+}$ and $\rho_{-}$ in front of $|x_{1}-x_{2}|^{-2[\psi]+1}$ are the $U(1)_{V}\times U(1)_{A}$ charge densities.
In the limit of half-integer dimension, $2[\psi]\in\NN^{+}$, it results
\begin{align}
\nonumber
& [\psi^*(t,x_1),\psi(t,x_2)]_{\k}=\sin(\frac{\pi\k}{2})\sum_{n=0}^{2[\psi]-1}\frac{1}{n!}\delta^{(n)}(x_1-x_{2})\mathcal{O}_{n}(x_2)\\
\label{OPEtheta2}
& =\sin(\frac{\pi\k}{2})\left\{-\ri\frac{1}{(2[\psi]-1)!}\delta^{(2[\psi]-1)}(x_{1}-x_{2})-
 [\rho_{+}(x_2)+\rho_{-}(x_2)]
\frac{1}{(2[\psi]-2)!}\delta^{(2[\psi]-2)}(x_{1}-x_{2})+
\ldots\right\}\,.
\end{align}
The expressions (\ref{OPEtheta}) and (\ref{OPEtheta2}) give the $\k$-commutator at equal time for massless (gapless)
anyons of dimension $[\psi]$ and statistical parameter $\k$.  
They are in sharp contrast with the non-relativistic expression for massive anyons \cite{Goldin:2003kj} where 
only  $\delta(x_1-x_2)$ appears on the right-hand side. 
However, from a dimensional analysis perspective, this difference is expected since in the massless case the scaling dimension of $\psi$ can differ from $1/2$. On the contrary, non-relativistic massive anyons always have scaling dimension $1/2$ and therefore the $\delta$-function is recovered in the $\k$-commutator \cite{Goldin:2003kj}.
 For instance, in the anyonic Thirring model discussed in section (\ref{sect-Thirring}) we have 
\begin{equation}
[\psi]=\frac{\k}{2}+\frac{g^2}{\k+2g}
\end{equation} 
where $g$ is the coupling constant. Hence, the $\k$-commutator in the Thirring model is equal to $\delta(x_1-x_2)$ only for special couplings
\begin{equation}
g_{\pm}=\frac{1}{2}\left(1-\k\pm\sqrt{1-\k^2}\right)\,.
\end{equation}

Finally, we end this appendix commenting on the limit $\k\rightarrow0$ for free massless anyons. 
This corresponds to take the limit to free massless bosons with  vanishing scaling dimension, $[\psi]=0$. 
Then, no $\delta$-function can appear in the $\k$-commutator. Indeed, the operator product expansion (\ref{OPEtheta}) gives
\begin{equation}
[\psi^*(t,x_1),\psi(t,x_2)]_{\k\rightarrow 0}=\langle [\psi^*(t,x_1),\psi(t,x_2)]_{0}\rangle=0\,.
\end{equation}

\end{document}